\documentclass[floatfix,twocolumn,showpacs,prb,amsmath,amssymb]{revtex4} 
\usepackage{graphicx} 

\newcommand{\sps}[1]{$^{\mathrm {#1}}$}

\newcommand{\sub}[1]{$_{\mathrm {#1}}$}
\newcommand{\subm}[1]{_{\mathrm {#1}}}

\newcommand{\cro}[1]{{#1}Cu\sub{3}Ru\sub{4}O\sub{12}}
\newcommand{\acro}{\cro{\textit{A}}}
\newcommand{\ncro}{\cro{Na}}
\newcommand{\ccro}{\cro{Ca}}
\newcommand{\lcro}{\cro{La}}
\newcommand{\sro}{Sr\sub{2}RuO\sub{4}}
\newcommand{\pio}{Pr\sub{2}Ir\sub{2}O\sub{7}}
\newcommand{\lvo}{LiV\sub{2}O\sub{4}}

\newcommand{\tc}{$T_{c}$}
\newcommand{\EF}{$E\subm{F}$}
\newcommand{\etal}{\textit{et~al.}}

\newcommand{\CuKa}{Cu$K_{\alpha1}$}

\begin{document}

\title{Suppression of the Mass Enhancement in CaCu$_3$Ru$_4$O$_{12}$}

\author{Soutarou~Tanaka}
\affiliation{Department of Physics, Graduate School of Science, 
Kyoto University, Kyoto, 606-8502, Japan}

\author{Hiroshi~Takatsu}
\affiliation{Department of Physics, Graduate School of Science, 
Kyoto University, Kyoto, 606-8502, Japan}

\author{Shingo~Yonezawa}
\affiliation{Department of Physics, Graduate School of Science, 
Kyoto University, Kyoto, 606-8502, Japan}

\author{Yoshiteru~Maeno}
\affiliation{Department of Physics, Graduate School of Science, 
Kyoto University, Kyoto, 606-8502, Japan}

\email{maeno@scphys.kyoto-u.ac.jp}

\date{\today}


\begin{abstract}
We have investigated heavy-fermion behavior of the transition-metal oxides
{\acro}
(\textit{A}~=~Na, Ca, La, and their mixtures).
It has been known that
{\ccro} exhibits Kondo-like behavior attributable to Cu\sps{2+} 3\textit{d} electrons,
similar to that of some Ce-based heavy-fermion systems.
However, we find striking \textit{suppression} of the mass enhancement in {\ccro},
in which the Kondo-like effect is most pronounced.
Such decrease of the density of states is reminiscent of the coherent gap formation
in Kondo lattice systems. 
Nevertheless, the behavior can not be interpreted within the conventional Kondo picture 
with localized moments, because the Cu electrons are apparently itinerant.
The present results indicate the importance of the duality of localized and itinerant nature,
found also in some other \textit{d}-electron systems which exhibit the Kondo-like behavior.

\end{abstract}


\maketitle

\section{Introduction}

Since the discovery of high-{\tc} superconductivity 
in the cuprates (La,Ba)\sub{2}CuO\sub{4}~\cite{Bednorz}
and
spin-triplet superconductivity in the ruthenate {\sro},~\cite{214}
considerable attention has been paid to
electronic properties of perovskite-related transition-metal oxides.
In some of these oxides electron correlations play important roles.
Cupro-ruthenates {\acro} (\textit{A} = Na, Ca, La)
\cite{%
Ebb2002JSolStaChem,%
Rami2004SolStaCommun,%
Koba2004JPSJ,%
Tran2006PRB,%
Xiang2007PRB,%
Krimmel2008PRB,%
Tanaka2009JPSJ,%
Sudayama2009%
}
are transition-metal oxides with a perovskite-related structure
and have recently been widely studied because of
their relatively large electronic specific heat coefficients
$\gamma$~=~70-140~mJ/(f.u.{\,}mol{\,}K\sps{2})
(f.u.:{\,}formula unit), as well as
their metallic conductivity originating from the \textit{d} electrons.
The values of $\gamma$ of {\acro} are much larger
than $\gamma$~$\simeq$~6~mJ/(mol{\,}K\sps{2}) of
the metallic transition-metal oxide RuO\sub{2}
without strong electron correlations,~\cite{passen}
but comparable to $\gamma$~=~38~mJ/(mol{\,}K\sps{2}) of
the strongly correlated metal {\sro}.~\cite{214}
The large values of $\gamma$ indicate
a realization of a Fermi liquid state with a heavy effective mass,
which is rare among \textit{d}-electron systems.

More interestingly, only for {\ccro}
with the formal valence of 2+ for Cu, 
similar to the insulating parent compounds of high-{\tc} cuprates,
Kondo-like behavior has been reported.~\cite{Koba2004JPSJ,Tanaka2009JPSJ,Sudayama2009}
A broad peak at around 200~K in the magnetic susceptibility $\chi(T)$
was found by Kobayashi~\textit{et~al.},~\cite{Koba2004JPSJ}
reminiscent of
those for some Ce-based heavy-fermion compounds with high Kondo temperatures
(valence fluctuation systems).~\cite{Kishimoto}
Kobayashi {\etal} regarded the peak in $\chi(T)$ of {\ccro}
as possible evidence of the lattice Kondo effect
between the localized Cu\sps{2+} electrons ($S=1/2$, 3\textit{d}\sps{9})
and the itinerant electrons originating from the Ru~4\textit{d} orbitals.
However, more recent experiments revealed profound duality of 
localized and itinerant characters of the Cu electrons.
First, for the origin of the mass enhancement, 
we recently clarified that electron correlations among the itinerant Ru electrons 
should be dominant and the Kondo-like effect provides, 
if any, only a minor contribution.~\cite{Tanaka2009JPSJ} 
Moreover, recent copper core-level X-ray photoemission spectroscopy (XPS) 
at room temperature by Sudayama~{\etal}.~\cite{Sudayama2009} 
revealed that {\ccro} shows the largest Cu DOS around {\EF} among {\acro}, 
indicating that the Cu electrons are in fact itinerant. 
Therefore, the conventional Kondo picture with localized electrons in a metal 
clearly fails to account for the Kondo-like behavior with enhanced mass in {\ccro}. 
Nevertheless, the Kondo-like picture is supported by
high-resolution photoemission spectroscopy (PES) measurements, 
which revealed that a prominent peak in the density of states (DOS) 
exists just at the Fermi level {\EF} and  grows below 100~K.~\cite{Sudayama2009}  
For the clarification of the duality of the Cu electrons in {\ccro}, 
namely the presence of the Kondo-like resonant state without localized moments, 
a detailed comparison with the other members of {\acro}, 
which exhibit the enhanced mass without Kondo-like behavior, 
is a promising experimental approach.

For \textit{f}-electron-based heavy-fermion systems,
the lattice Kondo effect is widely accepted as the dominant origin of the heavy mass.
Among metallic oxides with \textit{f} electrons,
the pyrochlore iridate {\pio} with localized Pr moments also exhibits
both the lattice Kondo effect and the mass enhancement~\cite{Nakatsuji2006PRL}.
In contrast with the {f}-electron-based systems, 
a general picture is still lacking to explain  the Kondo-like behavior 
found in a few metallic \textit{d}-electron systems 
with strong electron correlations.~\cite{hftr}
In these systems, the duality of localized and itinerant characters
appears to be an important common feature.
For example, the spinel vanadate {\lvo}
with local magnetic moments as well as itinerant electrons of vanadium 
exhibits physical properties similar to those of heavy-fermion systems.~
\cite{Kondo1997PRL}
For this compound, the importance of hybridization 
between two distinct orbitals~\cite{ani1999PRL}
and the closeness to orbital selective Mott transition~\cite{arita}
have been emphasized as the origin of the enhanced mass.
However, alternative explanations are also given based on the geometrical frustration, 
which suppresses long-range order and leads to fluctuations giving Kondo-like behavior.~
\cite{Lacroix2001}
Another example is the ruthenate Ca\sub{1.8}Sr\sub{0.2}RuO\sub4 
on the verge of Mott transition to an insulator 
with antiferromagnetic order of ruthenium local moments.
In high-resolution angle-resolved PES, quasiparticle peaks that grow 
at low temperatures have been observed.~\cite{simo2009PRL}
In {\ccro}, the additional presence of the itinerant heavy-mass Ru electrons allows us to control 
the duality of the Cu electrons which is responsible to the Kondo-like behavior. 
Thus, {\ccro} may provide an important clue to gain a general picture of the Kondo-like behavior 
found in some \textit{d}-electron systems.

In this paper, in order to examine the role of the Kondo-like effect on the mass enhancement of {\ccro},
we compare the magnetic susceptibility
and electronic specific heat of
not only the end-members {\ncro}, {\ccro} and {\lcro},
but also
the solid solutions {\cro{(Na,Ca,La)}},
in which 
Na\sps{+} and La\sps{3+} ions are partially substituted for Ca\sps{2+} in {\ccro}.
We use the Cu formal valence in this paper although the Cu valence may not be well-defined.
While $\chi(T)$ suggests the presence of a Kondo-like effect
only in compositions close to {\ccro},
we have revealed an additional, apparently \textit{negative} effect on the mass enhancement.
Such decrease of the DOS can not be interpreted within the conventional Kondo scenario
and suggests a possible formation of a gap structure in the electronic DOS at low temperatures.

\section{Experimental}
\begin{figure}[h!]
\includegraphics[width=0.42\textwidth]{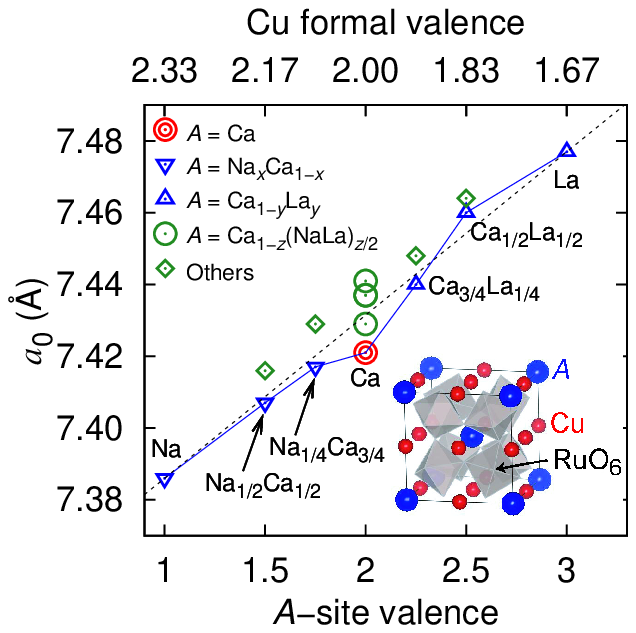}
\caption{
	(color online)
	Variation of the cubic lattice parameter $a_0$
	against the average valence of \textit{A} ions.
	The small blue triangles
	represent
	\cro{Na$_{x}$Ca$_{1-x}$}
	and
	\cro{Ca$_{1-y}$La$_{y}$}
	(valence series).
	The large green circles
	represent
	\cro{Ca$_{1-z}$(NaLa)$_{z/2}$}
	(charge-disorder series).
	The broken line connects the data points for A~=~Na and La for a guide to the eyes.
	Inset: Crystal structure of {\acro},
	generated using the program ``VESTA''%
	.~\cite{VESTA200806}
	The corner-sharing octahedra represent RuO\sub{6},
	in which the Ru ion is located at the center and the O ions occupy the corners.
	\label{fig:a0V}
	}
\end{figure}
Polycrystalline samples of \cro{(Na,Ca,La)} were prepared
by a solid state reaction as described in Ref.~\cite{Tanaka2009JPSJ}
Powder X-ray diffractometry with {\CuKa} radiation at room temperature indicates that
the samples were almost single-phased, containing at most a few percent of CuO and RuO\sub{2} (not shown).
Their structure, shown in the inset in Fig.~\ref{fig:a0V}, 
is cubic with a calculated lattice parameter $a_0$ at room temperature
that increases
from $7.386\pm0.001$~{\AA} for \textit{A}~=~Na to $7.477\pm0.001$~{\AA} for \textit{A}~=~La
as shown in Fig.~\ref{fig:a0V}.
Among the samples in this Letter, we will focus on the following two groups.
The members with \textit{A}~=~Na$_{x}$Ca$_{1-x}$ and Ca$_{1-y}$La$_{y}$
are referred to as the valence series,
where the formal valence of Cu varies,
and the members with \textit{A}~=~Ca$_{1-z}$(NaLa)$_{z/2}$
are referred to as the charge-disorder series,
where the average valence of \textit{A} ions is kept at two,
corresponding to the Cu formal valence of two.

\section{Results}
\subsection{Susceptibility}
\begin{figure*}[t]
\includegraphics[width=0.87\textwidth]{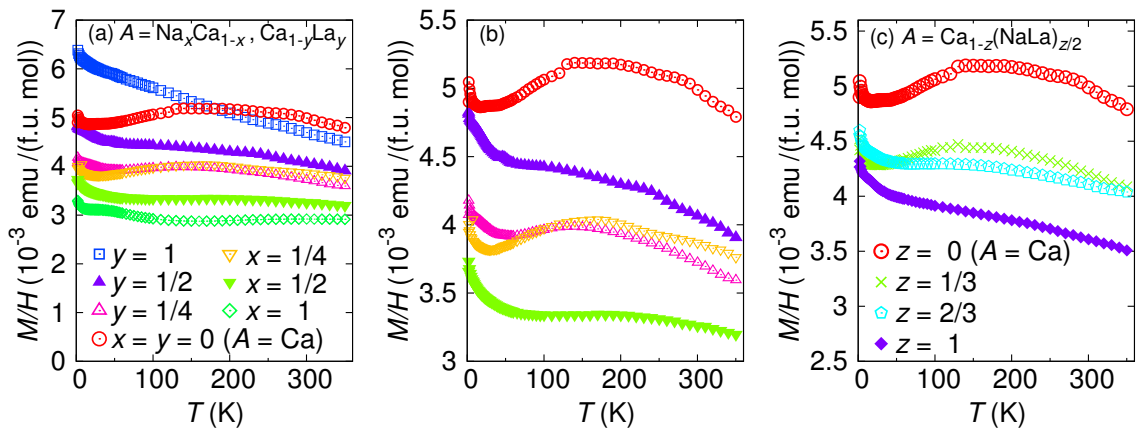}
\caption{
	(color online)
	Temperature dependence of DC magnetic susceptibility $\chi(T)=M(T)/H$ under 10~kOe
	for the valence series (a),
	for the valence series plotted in an enlarged vertical scale (b),
	and for charge-disorder series (c).
	\label{fig:chi}
}
\end{figure*}

Figure~\ref{fig:chi} displays their DC susceptibility $\chi(T)=M(T)/H$ 
measured at 10~kOe from 1.8~K to 350~K 
with a commercial SQUID magnetometer (Quantum Design, model MPMS).
First, we compare $\chi(T)$ for the valence series.
As shown in Fig.~\ref{fig:chi}(a),
a broad peak at around 180~K is observed for {\ccro},
consistent with earlier reports,~\cite{Koba2004JPSJ,Tanaka2009JPSJ}
relating the temperature dependence with a Kondo-like effect.
As $x$ or $y$ increases, the peak is gradually obscured as evident in Fig.~\ref{fig:chi}(b) 
with an enlarged vertical scale for \textit{A}~=~Na\sub{1/4}Ca\sub{3/4}, Ca, and Ca\sub{3/4}La\sub{1/4}.
The clear peak suggests a peculiarity of {\ccro} with the Cu formal valence of two.
Next, we examine $\chi(T)$ for the charge-disorder series with the Cu formal valence of two
in Fig.~\ref{fig:chi}(c).
As $z$ increases, it is expected that the local charge disorder at the \textit{A} site increases.
It is clear that the peak disappears for large $z$,
indicating that the charge disorder disturbs the peculiarity of {\ccro}.

\subsection{Specific heat}
\begin{figure*}[t]
\includegraphics[width=0.9\textwidth]{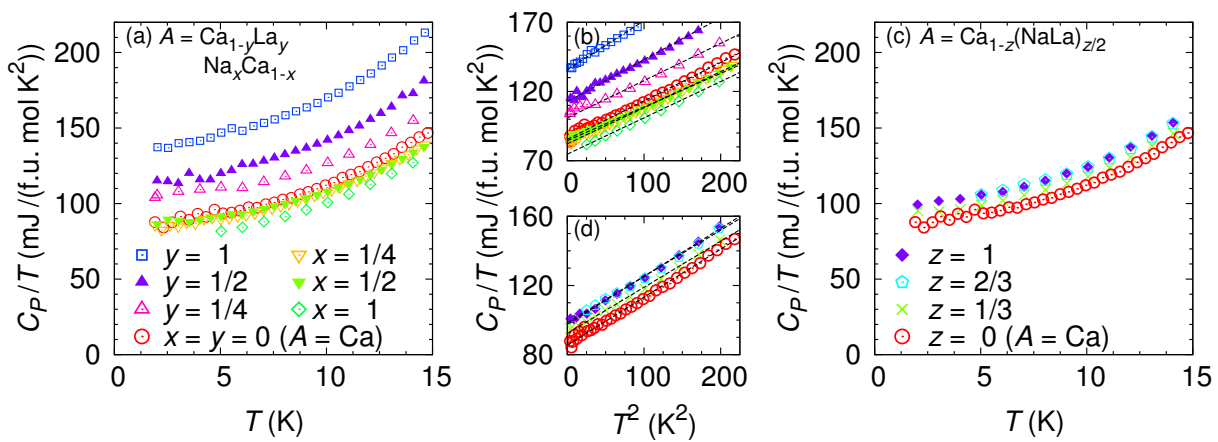}
\caption{
	(color online)
	Electronic specific heat divided by temperature $C_P(T)/T$.
	(a) $C_P(T)/T$ vs $T$ and 
	(b) $C_P(T)/T$ vs $T^2$ for the valence series.
	(c) $C_P(T)/T$ vs $T$ and 
	(d) $C_P(T)/T$ vs $T^2$ for the charge-disorder series.
	The number of displayed data points are reduced to avoid overlap in the plots for a clear presentation.
	The dashed lines in (b) and (d) are obtained by the fitting the equation
	$C_P(T)/T={\gamma}+{\beta}T^2$ to the data from 5~K to 15~K.
	\label{fig:cp}
}
\end{figure*}

Figure~\ref{fig:cp} displays their specific heat $C_P(T)$ at low temperatures
measured with a commercial calorimeter (Quantum Design, model PPMS).
The relation $C_P(T)/T={\gamma}+{\beta}T^2$ holds from 5~K to 15~K for all the samples,
yielding the electronic specific heat coefficients ${\gamma}$
consistent with our earlier report.~\cite{Tanaka2009JPSJ}
The values of $\gamma$ of all the present samples are relatively large.
Here, let us focus on the valence series and the charge-disorder series separately.
In the valence series, $\gamma$ increases
as the \textit{A} ions vary from Na\sps{+} to Ca\sps{2+} to La\sps{3+}
as shown in Figs.~\ref{fig:cp}(a) and \ref{fig:cp}(b).
In the charge-disorder series, interestingly, 
the samples with less disorder
have smaller values of $\gamma$
as shown in Figs.~\ref{fig:cp}(c) and \ref{fig:cp}(d).

\begin{figure}[t]
\includegraphics[width=0.42\textwidth]{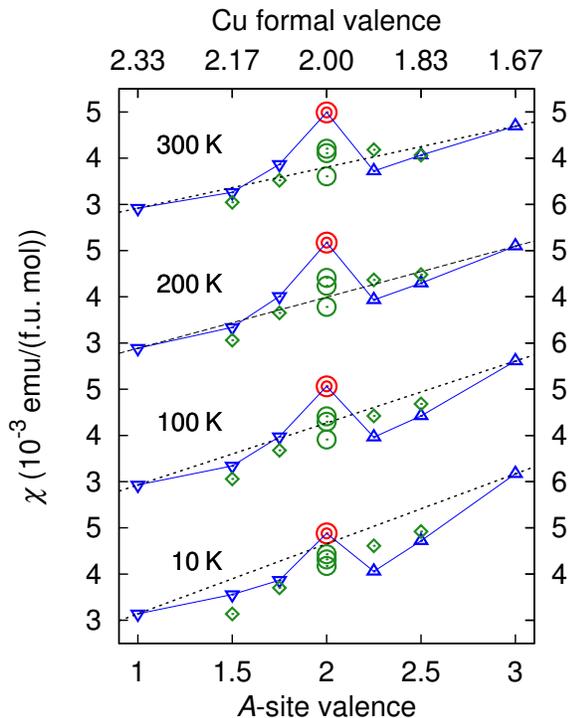}
\caption{
	(color online)
	Variation of the susceptibility $\chi$ at various temperatures
	against the average valence of the \textit{A}-site ions in {\acro}.
	\label{fig:chi-V}
}
\end{figure}

\begin{figure}[t]
\includegraphics[width=0.42\textwidth]{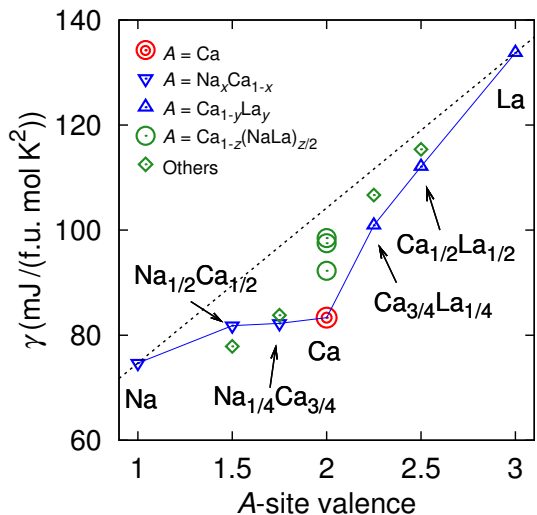}
\caption{
	(color online)
	Variation of the electronic specific heat coefficient $\gamma$
	against the average valence of the \textit{A}-site ions.
	\label{fig:gamma-V}
}
\end{figure}

In order to clarify these tendencies, the variations of 
both the magnetic susceptibility $\chi$
and the electronic specific heat coefficient $\gamma$
as functions of the average valence of \textit{A} are plotted 
in Figs.~\ref{fig:chi-V} and \ref{fig:gamma-V}.
There is a general tendency of $\chi$ and $\gamma$ to increase with the \textit{A}-site valence.
We note that the lattice parameter also exhibits a systematic increase, as shown in Fig.~\ref{fig:a0V}. 
Since this variation in the lattice parameter is known to cause little change 
in the bond angles of Ru-O-Ru as well as of Cu-O-Ru,~\cite{Ebb2002JSolStaChem} 
it is perhaps not the main origin of the change in $\chi$ and $\gamma$, 
as already argued by Ramirez \textit{et~al.}~\cite{Rami2004SolStaCommun} 
Rather, the change in the number of Cu and Ru electrons filling the states near the Fermi level 
is considered as the main origin for the change in the DOS.~\cite{Tanaka2009JPSJ} 

It is evident from Figs.~\ref{fig:chi-V} and \ref{fig:gamma-V} that 
{\ccro} exhibits peculiar properties among the {\acro} systems. 
As a function of the \textit{A}-site valence, 
$\chi$ shows a distinct peak below room temperature, 
although at low temperatures the peak becomes weaker. 
Because of the peak, one may expect a larger DOS and thus larger $\gamma$ for \textit{A}~=~Ca.
Contrary to the expectation, $\gamma$ exhibits a \textit{dip}, 
rather than a peak, at \textit{A}~=~Ca. 
The peak in $\chi$ as well as the dip in $\gamma$ is readily obscured
by introducing charge disorder at the \textit{A}-site,
indicating the importance of a well-ordered lattice.
These peak and dip lead to a large Wilson ratio of 4.0 for {\ccro},~\cite{Tanaka2009JPSJ}
indicative of enhanced magnetic fluctuations.


\section{Discussion and Conclusion}
As we have previously reported,~\cite{Tanaka2009JPSJ}
the large values of $\gamma$ of all \cro{(Na,Ca,La)}
suggest that the mass enhancement is mainly ascribable to correlations among itinerant Ru electrons.
This is the major premise in the following discussions.

Since {\ccro} exhibits Kondo-like behavior,
$\gamma$ is expected to be enhanced at low temperatures
in view of the conventional Kondo effect.
In contrast, the dip is observed in $\gamma$ for {\ccro}, 
for which the formal valence of Cu is exactly two. 
In analogy with the insulating parent phase of the high-{\tc} cuprates,
it may seem that the reduction of the Cu DOS at {\EF} occurs 
due to the localization of the Cu\sps{2+} electrons for {\ccro}. 
Such localization may overcome the enhancement of the DOS due to the Kondo-like effect.

However, the large value of $\chi$ for {\ccro} suggests that the Cu DOS is actually higher 
for this compound even at room temperature,
in contrast to the expectation for the localized Cu\sps{2+} states. 
In fact, the Cu core-level XPS indicates 
that the Cu DOS at {\EF} is higher for {\ccro} among the {\acro} compounds 
even at room temperature.~\cite{Sudayama2009}

Thus, an alternative interpretation is required to explain the suppression in $\gamma$ for {\ccro}, 
compatible with the enhanced susceptibility as well as 
the optical evidence for the enhanced Cu DOS at {\EF}.
In heavy-fermion systems,
it has been shown both experimentally and theoretically~\cite{ikeda1996,Shimada2002}
that a gap structure is formed from hybridization 
between localized \textit{f}-electrons and conduction electrons at half-filling,
leading to the Kondo semiconductor state.
We speculate that a similar hybridization gap is formed at low temperatures.~\cite{Sudayama2009} 
In {\ccro}, a metallic state with a substantial DOS at {\EF} is maintained
because of the multiband nature of the electronic states originating from the Ru orbitals.
The observed reduction of the peak in $\chi$ and the recovery of $\gamma$ 
for the charge-disorder series shown in Figs.~\ref{fig:chi-V} and \ref{fig:gamma-V} 
are consistent with the coherence nature required in such a hybridization gap.

In summary,
we have systematically investigated unusual electronic states 
in a series of transition-metal oxides {\acro} (\textit{A} = Na, Ca, La and their mixtures).
While all of these materials exhibit heavy-mass behavior,
a striking \textit{suppression} of the mass enhancement occurs in compositions close to {\ccro}, 
although the Kondo-like effect is most pronounced in {\ccro}.
We argued that a many-body effect in the presence of the enhanced Cu DOS,
in addition to the correlated bands of Ru~4\textit{d} electrons, 
leads to a phenomenon of reduction in DOS at {\EF} at low temperatures.
We speculate that a formation of a gap structure just at {\EF} is responsible for this reduction.
Such development of the electronic structure at low temperatures
is different from the conventional lattice Kondo effect in \textit{f}-electron systems
with clear localized moments. 
The distinct electronic state of {\ccro} among the {\acro} compounds,
including the striking suppression of the DOS, 
demonstrated the importance of the duality of localized and itinerant nature of the \textit{d} electrons. 
The results of this investigation should provide a useful viewpoint in clarifying the mechanisms of
the mass enhancement and Kondo-like effect found in some other \textit{d} electron systems. 

\acknowledgements
We would like to thank Takashi Mizokawa, Takaaki Sudayama, Norio Kawakami, and Yuji Matsuda 
for valuable discussions.
This work was supported by a Grant-in-Aid
for the Global COE Program ``The Next Generation of Physics, Spun from Universality and Emergence''
from the Ministry of Education, Culture, Sports, Science and Technology (MEXT) of Japan.
It was also supported by Grants-in-Aid for Scientific Research from MEXT
and from the Japan Society for the Promotion of Science.

\bibliography{ACRO_PRB_090608}

\end{document}